\documentclass[12pt]{article}
\usepackage{epsfig,fullpage}
\newcommand{\lapprox}{\raisebox{-0.5ex}{$\
\stackrel{\textstyle<}{\textstyle\sim}\ $}}
\newcommand{\gapprox}{\raisebox{-0.5ex}{$\
\stackrel{\textstyle>}{\textstyle\sim}\ $}}
\newcommand{\One}{1\kern-4.5pt1}
\begin{document}
\begin{center}

\begin{flushright}
     SWAT/02/327 \\
January 2002
\end{flushright}
\par \vskip 10mm

\vskip 1.2in

\begin{center}
{\LARGE\bf
Mesonic wavefunctions in the 
\vskip 0.12in
three-dimensional  
Gross-Neveu model}

\vskip 0.7in
S.J. Hands $\!^a$, J.B. Kogut $\! ^b$ and
 C.G. Strouthos $\!^{a}$ \\
\vskip 0.2in

$^a\,${\it Department of Physics, University of Wales Swansea,\\
Singleton Park, Swansea, SA2 8PP, U.K.} \\
$^b\,${\it Department of Physics, University of Illinois at Urbana-Champaign,\\
Urbana, Illinois 61801-3080, U.S.A.}\\
\end{center}

\vskip 1.0in 
{\large\bf Abstract}
\end{center}
\noindent
We present results from a numerical study of bound state 
wavefunctions in the (2+1)-dimensional 
Gross-Neveu model with staggered lattice fermions
at both zero and nonzero temperature.
Mesonic channels with varying quantum numbers are identified and analysed.
In the strongly coupled chirally broken phase at $T=0$ the wavefunctions
expose effects due to varying the interaction strength more effectively than 
straightforward spectroscopy. In the weakly coupled chirally restored phase
information on fermion -- anti-fermion scattering is recovered. In the hot
chirally restored phase we find evidence for a screened interaction. The 
$T=0$ chirally symmetric phase is most readily distinguished from the 
symmetric phase at high $T$ via the fermion dispersion relation.

\newpage

\section{Introduction}

Since the development of lattice gauge theory, many efforts have been
made to measure the mass spectrum of 
hadrons.  The spatial structure of 
hadrons, on the other hand, has not been as well-studied numerically,
although experimentalists have measured quantities like
the charge radius of the pion \cite{Amendolia84_Measurement_Pion}.
Numerical investigations of hadronic wavefunctions
were initiated in 1985 by Velikson and Weingarten \cite{wvfn1} 
and immediately afterwards they were 
followed by various other groups \cite{wvfn2,degrand}.
Knowing the form factor of the hadrons would allow one to calculate
their radii and higher moments. 
Hadronic wavefunctions are also useful in the calculation of masses
because they can be used as 
trial wavefunctions in the construction of hadronic operators
with improved overlap on the  
ground state at small Euclidean time separation.
Furthermore, at high
temperature DeTar pointed out the possibility of ``confinement'' 
at large distance
scales \cite{DeTar}.
Mesonic wavefunctions  at non-zero temperature were investigated recently on an
anisotropic 
lattice in the quenched approximation by the QCD-TARO Collaboration \cite{taro}.
Their results suggest that above $T_c$ there can be low energy excitations 
in the mesonic channels that are metastable bound states 
i.e. the quark and anti-quark
tend to stay together at least for Euclidean time scales $ \sim 1/T$.
These quasiparticles would be characterized by a mass scale
given by the location of a peak
in the corresponding spectral function.
This picture is consistent with  mean field calculations
in the Nambu$-$Jona-Lasinio model \cite{hatsuda} which predict that the mesons 
associated with the chiral order parameter, namely the sigma
and the pions are ``soft modes'' just above $T_c$, i.e. their 
fluctuations acquire
a large strength with a small width in the spectral density, which implies 
the existence of long-lived quasiparticles.
It has also been shown that the spatial wavefunctions for various mesonic 
channels in the high temperature phase 
\cite {spatial_wvfn,Laermann} are strongly localized, because of the nontrivial magnetic interaction
of colour currents in the quark-gluon plasma.

In this paper we present results from a study of mesonic wavefunctions measured 
in numerical simulations of the 
$U(N_f)_V$-invariant three-dimensional Gross-Neveu model (GNM$_3$)  with both 
discrete $Z_2$ and continuous $U(1)$ chiral symmetries
at both zero and non-zero temperature. 
The model with $U(1)$ chiral symmetry is described by the 
following semi-bosonized Euclidean Lagrangian density:
\begin{equation}
{\cal L}= \bar{\psi}_i(\partial\hskip -.5em / + m_0 + \sigma + i \gamma_5 \pi)\psi_i
+ \frac{N_f}{2 g^{2}} (\sigma^{2}+ \pi^2).
\end{equation}
We treat $\psi_i$, $\bar{\psi}_i$  as four-component Dirac spinors and the index $i$ runs over $N_f$
fermion species.
In the case of a $Z_2$ chiral symmetry the $\pi$ fields are set to zero 
in the Lagrangian.
At tree level, the fields $\sigma$ and $\pi$ have no dynamics; they are truly 
auxiliary fields. However, they acquire dynamical content by dint of quantum effects
arising from integrating out the fermions. 
The model is renormalizable in the $1/N_f$ expansion unlike in the loop 
expansion \cite{rosen91}.
Apart from the obvious numerical advantages of working with a relatively simple model in a reduced
dimensionality
there are several other motivations for studying such a model: (i) at $T\!=\!0$ for sufficiently 
strong coupling $g^2>g_c^2$ it exhibits spontaneous chiral symmetry breaking 
implying dynamical mass generation for the fermion, 
the pion field $\pi$ 
being the associated Goldstone boson; (ii) the spectrum 
of excitations contains both baryons and mesons, 
i.e. the elementary fermions $f$ and
the composite $f\bar f$ states; 
(iii) the model has an interacting continuum limit
at the critical value of the coupling, which has a numerical value 
$g^2/a\approx1.0$ in the large-$N_f$ limit if a lattice
regularisation is employed \cite{kogut93}; 
(iv) numerical simulations of the 
model with chemical potential $\mu\not=0$ show qualitatively correct behaviour, 
unlike QCD
simulations \cite{hands95,pion}.

Let us briefly review the physical content of the model as predicted by the
large-$N_f$ approach \cite{rosen91,kogut93}. For $g^2>g_c^2$ the fermion 
has a dynamically generated mass $M_f$ equal, up to corrections of 
order $1/N_f$, to the scalar field expectation value
$\langle\sigma\rangle=g^2\langle\bar\psi\psi\rangle$. The inverse $M_f^{-1}$
defines a correlation length which diverges as $(g^2-g_c^2)^{-\nu}$
with the critical index $\nu=1+O(1/N_f)$. 
As a result
of $f\bar f$ loop corrections
the $\sigma$ and $\pi$ fields acquire non-trivial dynamics. In the case of the 
$U(1)$-symmetric model, the pseudoscalar $\pi$ couples to a Goldstone mode, and 
its propagator has a massless
pole in the chiral limit $m_0\to0$. 
For both $U(1)$ and $Z_2$ chiral symmetries 
the scalar $\sigma$, by contrast, has mass
$2M_f$ in the large-$N_f$ limit, and rather than exhibiting an isolated pole,
has a continuum of $f\bar f$ states extending all the way
down to this threshold, implying that if truly bound, its binding energy is 
$O(1/N_f)$ at best. To our knowledge there have been no analytic calculations of
the binding energy in this channel. Since all residual interactions 
are subleading in $1/N_f$, we surmise that all other mesons are
similarly weakly bound states of massive fermions, and hence effectively
described by a two-dimensional ``non-relativistic quark model''.
For $g^2<g_c^2$ the model 
is chirally 
symmetric, and hence all states massless as $m_0\to0$.
A dimensionful scale is still defined, however, by the width of a resonance in
$f\bar f$ scattering in the scalar channel; 
this diverges as $(g^2_c-g^2)^{-\nu}$
with the same exponent $\nu$. In this case we have no equivalent of the 
quark model to assist interpretation of the wavefunctions to be discussed below.

The large-$N_f$ approach \cite{general}
also predicts that for $g^2>g_c^2$, chiral symmetry is
restored as temperature $T$ is raised beyond $T_c=M_f/2\ln2$, or for
chemical potential $\mu>\mu_c=M_f$.
The phase diagram of GNM$_3$ with various global symmetries
at non-zero temperature and density
has been studied extensively in [12$-$17]. More specifically,
it has been shown that the thermally induced phase
transition of the $Z_2$-symmetric model belongs to the two-dimensional Ising universality class
and the  $U(1)$-symmetric model undergoes
a Berezinskii-Kosterlitz-Thouless transition \cite{bkt,babaev}
in accordance with the dimensional reduction scenario, which predicts
that the long-range behaviour at the chiral phase transition is that of the
$(d-1)$ spin model with the same symmetry, because the IR region of the
system is dominated by
the zero Matsubara mode of the bosonic field. An interesting aspect of 
GNM$_3$ is that it manifests both bulk and thermally-induced
chirally-symmetric phases, and one of the goals of the present study is to
compare and contrast their properties.

The shape and size of a hadronic state $M$ can be
observed through the equal time Bethe-Salpeter wavefunction given by: 
\begin{equation}
\Psi(\mathbf{x},t) = \int d{\mathbf{y}} \langle 0| 
\bar{\psi}(\mathbf{y},t) \psi(\mathbf{y}+\mathbf{x},t)|M \rangle.
\end{equation}
One can extract $\Psi$ from the correlation function $C(\mathbf{x},t)$
which is a convolution 
of the quark propagator $G_q$ and the anti-quark propagator $G_{\bar{q}}$
and is given by:
\begin{equation}
C(\mathbf{x},t) = \int d {\mathbf y} d {\mathbf y}_1 d {\mathbf y}_2
\langle 0| \Phi(\mathbf{y}_1) \Phi(\mathbf{y}_2) 
G_{q}(\mathbf{y},t; \mathbf{y}_1, 0) \Gamma
G_{\bar{q}}(\mathbf{y}+\mathbf{x},t; \mathbf{y}_2, 0) \Gamma |0 \rangle 
\label{correl_wvfn}
\end{equation}
The Dirac matrix $\Gamma$ selects the appropriate spin and parity quantum 
numbers for the meson, i.e., $\Gamma={\bf 1}$ for the spin zero scalar (S) and 
$\Gamma=\gamma_5$ for the spin zero pseudoscalar (PS).
$\Phi({\bf x})$ is an input trial wavefunction which is used as a source for the
construction of quark and anti-quark propagators. At large $t$ the contribution 
from the ground state dominates and 
\begin{equation}
C(\mathbf{x},t) \simeq \exp(-m_H t) \Psi(\mathbf{x}),
\label{wvfn_propor}
\end{equation}
where $m_H$ is the hadron mass.
As already mentioned in \cite{degrand} these wavefunctions are minimal Fock space wavefunctions,
because they do not overlap onto states for which the quark antiquark world line
has kinks crossing time $t$. 
Therefore, their use to calculate phenomenological numbers 
is an uncontrolled approximation.
However, as we shall see, they are 
convenient tools in lattice simulations to study the binding 
in the various mesonic channels.
Due to its simplicity, 
the GNM$_3$ can be studied at both  $T=0$ and $T \neq 0$  on lattices
with relatively large $L_t$, which is the main difficulty for QCD simulations.
In the next section we will discuss numerical 
results of mesonic masses and wavefunctions on large volumes
in both the scalar (S) and pseudoscalar (PS) channels. 
In our study we only measured the 
connected parts of the correlators. 
The noisy disconnected diagrams were neglected 
and we will discuss the implication of this in the next section.
We will also discuss the dependence of our results on the quark propagator 
source, on the 
lattice spatial extent $L_s$, the chiral 
symmetry group ($Z_2$ vs. $U(1)$) of GNM$_3$, the coupling 
$\beta \equiv 1/g^2$ and the number of fermion species in the model $N_f$. 
We will then 
present results 
in  the $T=0$ and  $T \neq 0$ symmetric phases.
The sigma and the pion are represented in the semi-bosonised 
GNM$_3$ action by bosonic auxiliary fields
and hence
the correlation functions in these channels 
including disconnected diagrams can be measured with 
relatively high statistics. Unfortunately, 
the same technique cannot be applied to the wavefunctions 
which are point-split quark four-point functions.
In another project, which in a sense is complementary to this one we are 
studying mesonic spectral functions including those of the auxiliary fields
using the
Maximum Entropy Method \cite{mem}.

\section{Simulations}
The fermionic part of the lattice action we have used for the semi-bosonized GNM$_3$
with $U(1)$ chiral symmetry is given by \cite{hands95}
\begin{eqnarray}
S_{fer} & = &   \bar{\chi}_i(x) M_{ijxy} \chi_j(y)   \nonumber \\ 
        & = &   \sum_{i=1}^{N} \left ( \sum_{x,y} \bar{\chi}_i(x) \mathcal{M}_{xy} \chi_i(y)
 +  \frac{1}{8} \sum_{x} \bar{\chi}_i(x) \chi_i(x)  
[ \sum_{ \langle \tilde{x},x \rangle}
\sigma(\tilde{x}) + i \epsilon(x) \sum_{ \langle \tilde{x},x \rangle}
\pi(\tilde{x}) ] \right ),
\end{eqnarray}
where $\chi_i$ and $\bar{\chi}_i$ are Grassmann-valued 
staggered fermion fields
defined on the lattice sites, 
the auxiliary fields $\sigma$ and $\pi$ are defined on the dual lattice
sites, 
and the symbol $\langle \tilde{x},x \rangle$ denotes the set of 8 dual lattice
sites $\tilde{x}$ surrounding the direct lattice site $x$.
The fermion kinetic operator $ \mathcal{M} $ is given by
\begin{equation}
\mathcal{M}_{xy} = 
\frac{1}{2} \sum_{\nu} \eta_{\nu}(x) \left[ \delta_{y,x+\hat{\nu}} -
\delta_{y,x-\hat{\nu}} \right]+m_0\delta_{xy},
\end{equation}
where $\eta_{\nu}(x)$ are the Kawamoto-Smit phases 
$(-1)^{x_0+\cdots+x_{\nu-1}}$,
and the symbol $\epsilon(x)$ denotes the alternating phase $(-1)^{x_0+x_1+x_2}$.
The auxiliary fields $\sigma$ and $\pi$ are weighted in the path integral by an 
additional factor corresponding to 
\begin{equation}
S_{aux}=\frac{N}{2g^{2}} \sum_{\tilde{x}} [\sigma^2(\tilde{x}) + \pi^2(\tilde{x})].
\end{equation}
The simulations were performed by using the 
standard hybrid Monte Carlo algorithm 
in which complex bosonic pseudofermion 
fields $\Phi$ are updated using the action 
$\Phi^{\dagger}(M^{\dagger}M)^{-1} \Phi$.
Unless stated otherwise, the bare fermion mass $m_0$ was set to zero.
According to the discussion in \cite{hands95}, simulation of $N$ staggered
fermions describes $N_f=4N$ continuum species; 
the full symmetry of the lattice model in the
continuum limit, however, is $U(N_f/2)_V \otimes U(N_f/2)_V \otimes U(1)$ 
rather than $U(N_f)_V \otimes U(1)$. 
At non-zero lattice spacing the symmetry group is smaller: $U(N_f/4)_V \otimes
U(N_f/4)_V \otimes U(1)$. 
In the  $Z_2$-symmetric model the $\pi$ fields are switched off and $M$ 
becomes real. In this case according to the discusion in \cite{kogut93} simulation of 
$N$ staggered fermions describes $N_f=2N$ continuum species.
Details about the algorithm and how we optimized 
its performance can be found in \cite{kogut93,hands95}.

Using point sources we calculated the zero momentum fermion correlator
at different values of the coupling $\beta$
and fitted to the function 
\begin{equation}
C_f(t)=A_f [e^{-M_ft}-(-1)^t e^{-M_f(L_t-t)}].
\end{equation}
The mesonic correlators are given by:
\begin{equation}
C_{M}(t)=\sum_{\mathbf{x},\bf{x}_1,\mathbf{x}_2} \Phi(\mathbf{x}_1) \Phi(\mathbf{x}_2)
W_{M}({\mathbf x}) G(\mathbf{x},t;\mathbf{x}_1,0)
G^{\dagger}(\mathbf{x},t;\mathbf{x}_2,0),
\end{equation}
where $W_{M}({\mathbf x})$ is a staggered fermion
phase factor which picks out a channel with particular
symmetry properties i.e. $W_{M}({\mathbf x}) = \epsilon(x)$ 
for the S channel and 
$W_{M}({\mathbf x}) = 1$ for the PS channel.
The function $\Phi(\mathbf{x})$ is either a point source
$\delta_{\mathbf{x},(0,0)}$ 
or a staggered fermion wall source $\sum_{m,n=0}^{L_s/2-1}
\delta_{\mathbf{x},(2m,2n)}$ \cite{gupta}.
In all the simulations we used point sinks.
These correlators were fitted to a function $C_{M}(t)$ given by
\begin{equation}
C_M(t)=A[ e^{-mt} + e^{-m(L_t-t)}] + \tilde{A} (-1)^t [e^{-\tilde{m}t} + e^{-\tilde{m}(L_t-t)}].
\end{equation}
The first square bracket represents the ``direct'' signal with mass $m$ 
and the second an ``alternating'' signal
with mass $\tilde{m}$.
Just as in four dimensions, composite operators made from staggered fermion
fields project onto more than one set of continuum quantum numbers.
To gain more insight one must transform to a basis with explicit spinor and
flavor indices \cite{burden}; the resulting bilinears 
$\bar q(\Gamma\otimes T)q$ in both S and PS channels, where $T$ acts on a
two-component flavor space, are summarised in
Table~\ref{tab:spinflav}.
\begin{table}[ht]
\setlength{\tabcolsep}{1.5pc}
\caption{Spin/flavor assignments of the fermion bilinears studied in the paper}
\label{tab:spinflav}
\begin{tabular*}{\textwidth}{@{}l@{\extracolsep{\fill}}ccccc}
\hline
& $(\Gamma\otimes T)_{dir}$& $J^P_{dir}$ & $(\Gamma\otimes T)_{alt}$&
$J^P_{alt}$\\
\hline
PS    &
$\gamma_5\otimes\One$ & $0^-$ & $-i\gamma_1\gamma_2\otimes\tau_3$ & $1^-$\\
S  &
$\One\otimes\One$ & $0^+$ & $i\gamma_0\gamma_3\otimes\tau_3$& $0^+$\\
\hline
\end{tabular*}
\end{table}
Also shown are the spin/parity quantum numbers $J^P$. 
In the direct channels these are the same as their four-dimensional 
equivalents, but the alternating channels are very different. In the PS channel
the alternating state is an anti-symmetric tensor, corresponding to spin-1,
whereas in the S channel the bilinear transforms trivially under rotations in
the 12 plane and is hence spin-0. The parity states are assigned by considering
transformation properties under 2+1 dimensional lattice parity:
\begin{eqnarray}
x=(x_0,x_1,x_2) &\mapsto& x^\prime=(x_0,1-x_1,x_2) \nonumber \\
\chi,\bar\chi(x)&\mapsto&(-1)^{x_1^\prime+x_2^\prime}\chi,\bar\chi(x^\prime)
\nonumber\\
\sigma(\tilde x)\mapsto\sigma(\tilde x^\prime) &;&
\vec\pi(\tilde x)\mapsto-\vec\pi(\tilde x^\prime).\label{eq:parity}
\end{eqnarray}
In continuum notation this corresponds to determining whether $\Gamma$ commutes
or anti-commutes with $\gamma_1\gamma_5$. It is important to note that unlike in
3+1 dimensions, the alternating channels do not contain ``parity partners'' of
the direct channels, or indeed necessarily even decribe states of the same
spin.

In a similar way the lattice  wavefunctions are given by: 
\begin{equation}
\Psi(\mathbf{x})=\sum_{\mathbf{x}^{\prime},\mathbf{x}_1,\mathbf{x}_2} \Phi(\mathbf{x}_1)
\Phi(\mathbf{x}_2) W_{M}({\mathbf x}) G(\mathbf{x}+\mathbf{x}^{\prime},t;\mathbf{x}_1,0)
G^{\dagger}(\mathbf{x}^{\prime},t;\mathbf{x}_2,0).
\label{eq:wav_latt}
\end{equation}
In the staggered formulation the easiest way to construct $\Psi$
is by computing it for spatial separations that are multiples of $2a$.
To facilitate comparison between the various
wavefunctions we normalize them to unity at zero separation.

\subsection{Zero Temperature Broken Phase}
In this section we discuss the numerical results for the  masses
and the wavefunctions extracted from the S and PS channels of GNM$_3$
at $T=0$. We generated statistics that vary from $700$ to $2000$ 
configurations, 
depending on how close to the critical 
point is the coupling for each simulation.
The average molecular dynamics trajectory length $\bar{\tau}\simeq1.2$. 

First, we measured the meson correlators $C_M(t)$ in the S and PS channels 
and the fermion correlator
$C_f(t)$ at different
values of the coupling $\beta$ from simulations of the $Z_2$-symmetric model.
An interesting observation is that the direct 
amplitude $A$ in the S correlator
is tiny compared to the alternating signal $\tilde{A}$, 
and appears to decrease rapidly as we decrease the coupling $\beta$,
whereas the PS correlator is dominated by the direct signal.  
For example, 
in the case of the S correlator for $\beta=0.75$ the fit parameters are  
$A=-0.00010(5)$, $\tilde{A}=250(10)$, $\tilde{m}=0.34(1)$, 
$M_f=0.175(1)$ and for
$\beta=0.80$ they are $A=-0.05(1)$, $\tilde{A}=235(5)$, $\tilde{m}=0.172(1)$, 
$M_f=0.0893(5)$. The bulk critical coupling is $\beta_c=0.85(1)$.
We infer that the alternating signal corresponds 
to a weakly bound state since its mass
is slightly less than $2M_f$. 
It appears that the flavor-singlet scalar channel S$_{dir}$
is completely dominated by disconnected diagrams in this phase,
and can only be studied using the auxiliary $\sigma$ field \cite{kogut93}.
Because of the smallness of $A$ in the S correlator  
($\tilde{A}$ in the PS correlator) for $\beta\lapprox\beta_c\simeq0.85$,
and the statistical fluctuations in our data
we do not quote values for their masses; they did, however, appear very light,
perhaps even consistent with zero, 
perhaps suggesting their
origin is due to the proximity of the chirally symmetric phase at
$\beta\gapprox0.85$.

Even in the case of continuous chiral symmetry, 
the meson extracted from the connected contribution in the PS$_{dir}$
channel has a mass
almost twice the fermion mass \cite{pion,mem} because it is {\sl not\/} the
physical pion:
the Goldstone mechanism in GNM$_3$ is fundamentally different from that
in QCD. In this model the disconnected diagrams are responsible for making the
pion light; numerically this can be accessed by studying correlators of the
auxiliary $\pi$ field \cite{pion,mem}.

\begin{figure}[p]

                \centerline{ \epsfysize=3.2in
                             \epsfbox{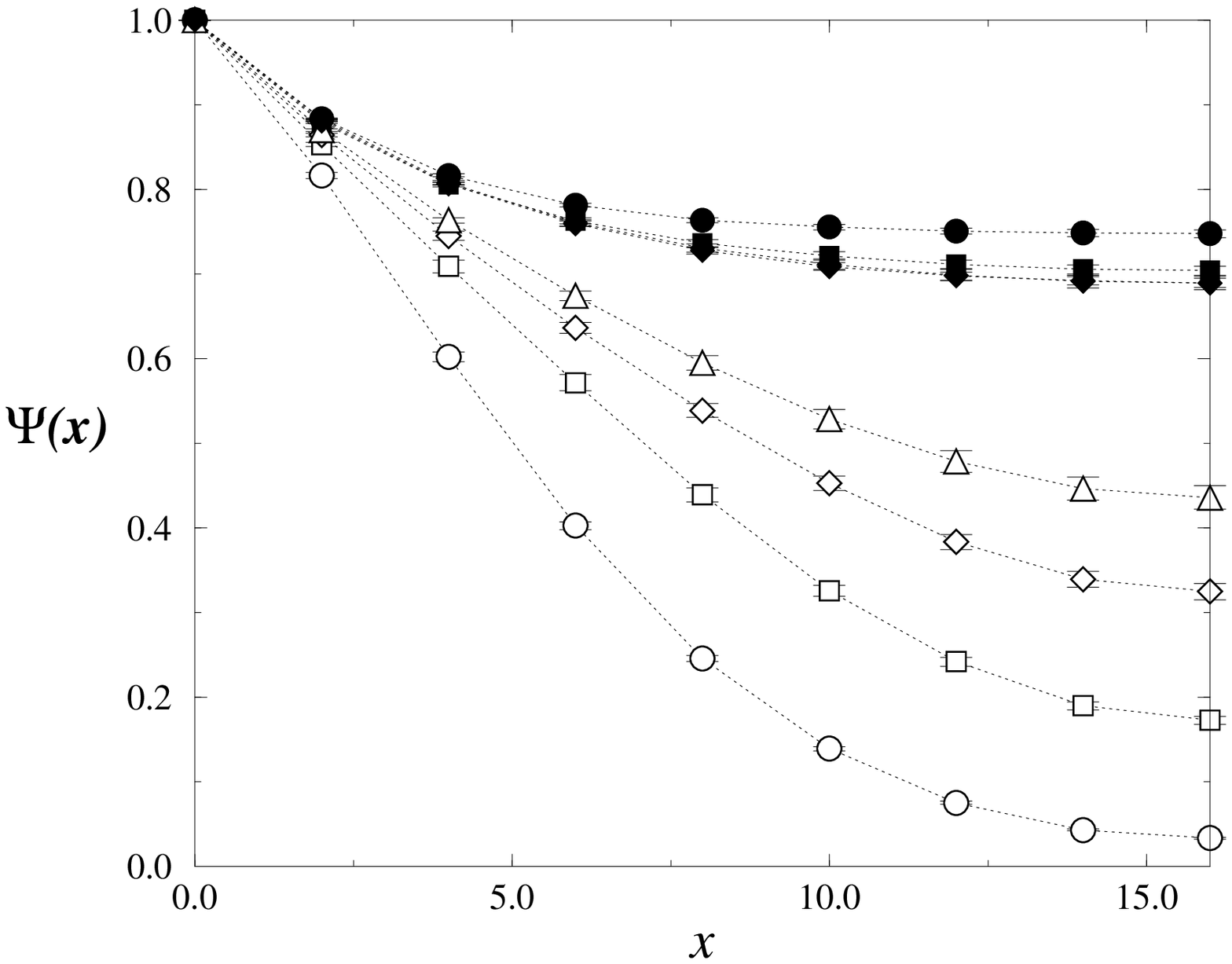}}

\smallskip
\caption[]{PS wavefunctions at $\beta=0.55$. Filled symbols correspond to wall source;
$t$ increases from top to bottom. Empty symbols correspond to point source; $t$
increases from bottom to top.}
\label{fig:w-p_source}
\end{figure}

In Fig.~\ref{fig:w-p_source} we present results for the PS channel 
wavefunctions of the $Z_2$ model  
with $N_f=4$ on 
a $32^2 \times 96$ 
lattice and coupling $\beta = 0.55$.
The wavefunctions were extracted from different timeslices ($t=11,21,31,41$) 
and were constructed from quark propagators with  both wall and point sources.
It is evident that the wavefunctions extracted from wall sources
stabilize at relatively small Euclidean times.
It is expected that the wall source
projects onto the ground state much more effectively than the point source,
because the wavefunction is very broad and it does not go to zero at 
the edge of the lattice, i.e. it resembles a wall wavefunction. 
The other advantage of the wall sources wavefunction is that 
it is constructed by convolving quarks with $\mathbf{p}=\mathbf{0}$,  
whereas the wavefunction extracted from point sources has more contamination 
from excited states at small times. 
As mentioned in the previous paragraph this weakly bound state has a mass which is very close to 
twice the fermion mass; in this case $M_{PS}=1.085(3)$ and $M_f=0.550(1)$. 
Because of the superior saturation properties, 
the various correlators that we will discuss in the next few paragraphs
were constructed from wall sources.

\begin{figure}[p]

                \centerline{ \epsfysize=3.2in
                             \epsfbox{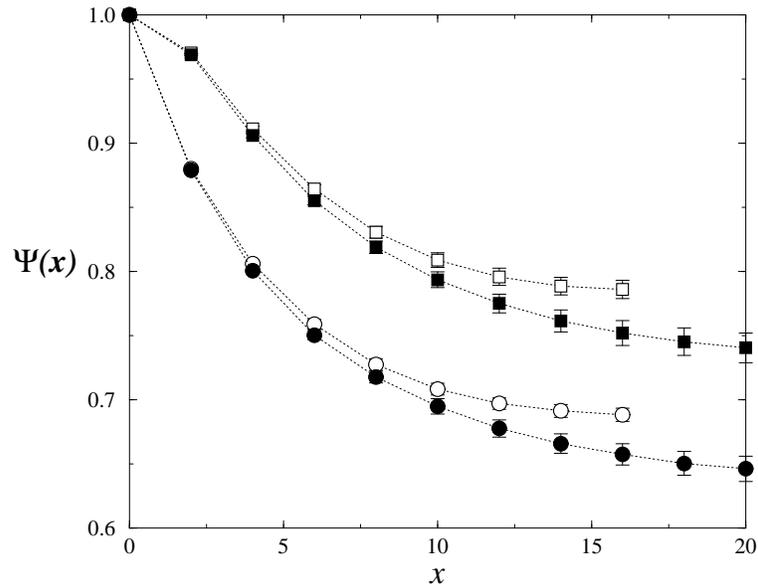}}

\smallskip
\caption[]{S and PS wavefunctions at $\beta=0.55$ and $L_s=32,48$.}
\label{fig:l=32_48}
\end{figure}

\begin{figure}[p]

                \centerline{ \epsfysize=3.2in
                             \epsfbox{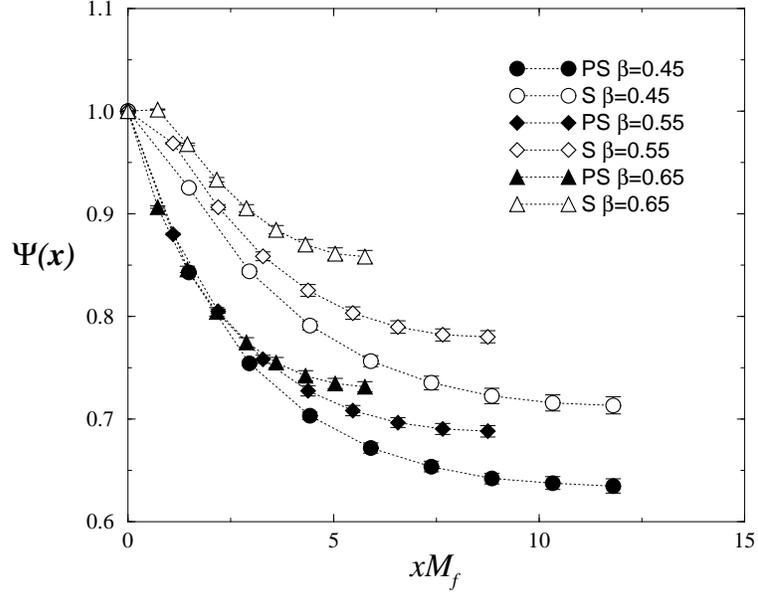}}

\smallskip
\caption[]{S and PS wavefunctions at $\beta=0.45,0.55,0.65$; the horizontal axis is rescaled by
the fermion mass $M_f$.}
\label{fig:rescaled_broken}
\end{figure}

In Fig.~\ref{fig:l=32_48} we present PS$_{dir}$ and S$_{alt}$
wavefunctions at coupling $\beta = 0.55$
and lattice sizes $32^2 \times 96$ and $48^2 \times 96$.  
The wavefunctions fill the lattice even when $L_s=48$ and it is 
obvious that finite size effects
are still large at the boundary. 
In Fig.~\ref{fig:rescaled_broken} we plot the $S_{alt}$ and
$PS_{alt}$ wavefunctions in the broken phase 
for $\beta=0.45,0.55,0.65$
with the horizontal axis rescaled by the fermion mass 
$M_f$ which is an inverse 
correlation length. 
For small distances (in physical units $xM_f\lapprox3$)
the three $PS_{dir}$ wavefunctions coincide, 
implying that in this regime the wavefunction
shape is that of the continuum limit theory. 
We interpret the deviations
at large distances as strong finite size effects; 
as implied by Fig.~\ref{fig:l=32_48},
the meson fills the lattice at these couplings.
The $S_{alt}$ wavefunctions which 
are broader than the $PS_{alt}$ wavefunctions 
deviate slightly from each other even near the origin,
indicative of stronger curvature in this region.

One clear effect is that the $f\bar f$ pair is much more closely
bound in the PS channel than in the S. We can explain this by adapting
the well-known argument from the non-relativistic quark model that 
a fermion -- anti-fermion system bound in an
$s$-wave is necessarily pseudoscalar due to the product of intrinsic parities
of the $f\bar f$ pair being negative. Some care must be taken in 2+1 dimensions;
whereas the $L=0$ spatial wavefunction is 
axially symmetric and hence positive parity, 
all higher angular momentum states, which 
have the form
$f(\vert \mathbf{r}\vert)e^{\pm iL\phi}$, 
where $f(r)\propto r^L$ for small $r$ and for non-singular potentials,
contain both positive and negative parity 
components (unlike in 3+1 dimensions where parity is given by $(-1)^L$),
implying that higher spin states fall into degenerate parity doublets if parity
is a good quantum 
number~\footnote{Intrinsic parity must also be defined with care:
recall that in 3+1 dimensions it is related to the spinor eigenstates of
$\gamma_0$. Since the continuum formulation of staggered fermions in 2+1
dimensions naturally results in 4-component spinors, we can effectively regard 
the parity transformation (\ref{eq:parity}) as a spatial inversion 
in 3 dimensions
followed by a rotation by $\pi$ radians in the 23 plane; since rotations are
continuous the 
$f$ and $\bar f$ spinor eigenstates maintain their respective intrinsic
parities under this sequence of transformations.}. 
The conclusion remains, however, that the only 
allowed $s$-wave state is the $0^-$, which naturally accounts for why the
PS$_{dir}$ wavefunction has the smallest extent. The behaviour of 
states with $L>0$ as a function of $\vert\mathbf{r}\vert$ 
might possibly explain the noticeable kink in the S wavefunction
at the smallest non-zero spatial separation; ideally
data nearer the continuum
limit would be needed to explore this more thoroughly.
 
\begin{figure}[p]

                \centerline{ \epsfysize=3.2in
                             \epsfbox{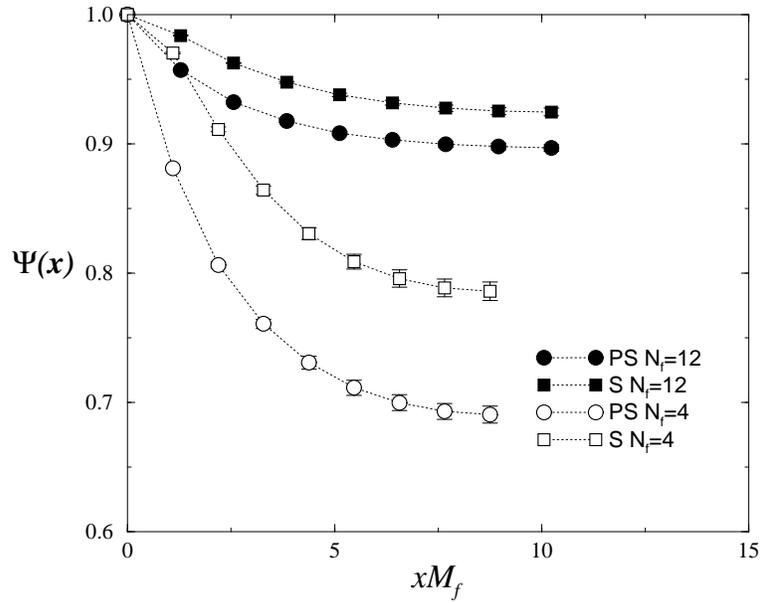}}

\smallskip
\caption[]{S and PS wavefunctions for $N_f=4$ and $N_f=12$; the horizontal axis is rescaled 
by $M_f$.}
\label{fig:n=4_12}
\end{figure}

In Fig.~\ref{fig:n=4_12} we present PS$_{dir}$ and
S$_{alt}$ wavefunctions measured in simulations 
with different numbers of fermion 
species in the system, i.e., $N_f=4,12$.  
The coupling is set to $\beta=0.55$ and the wavefunctions shown in this figure
were extracted from the $t=40$ timeslice. 
In order to compare the two cases and  
probe $1/N_f$ effects we 
rescaled the horizontal axis by multiplying the distance
with the fermion mass $M_f$. It is clear from the figure that 
as we switch on the 
interaction (i.e. decrease $N_f$) the binding increases 
and the wavefunction shrinks.
The ratio $M_{PS}/(2M_f)$ is $0.986(5)$ for $N_f=4$ and $0.999(1)$ for $N_f=12$.
These results imply that the wavefunctions in this model are far more sensitive to  $1/N_f$ effects 
than the masses or even the critical indices.
In this model the bulk critical
indices for finite $N_f$ are very close to the large-$N_f$ prediction 
and $1/N_f$ effects have been very difficult to measure even for 
relatively small 
$N_f$ \cite{largeN}.

\begin{figure}[p]

                \centerline{ \epsfysize=3.2in
                             \epsfbox{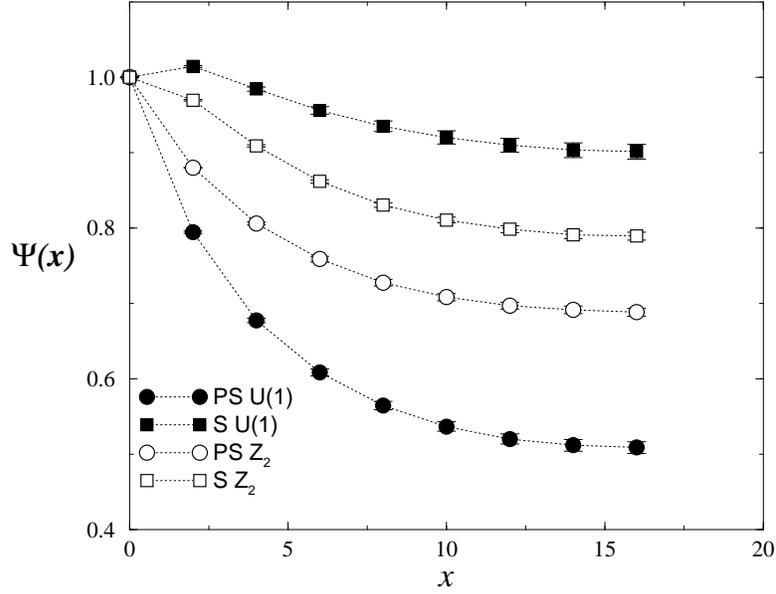}}

\smallskip
\caption[]{S and PS wavefunctions from the $U(1)$ and $Z_2$ GNM$_3$.}
\label{fig:z2_u1}
\end{figure}

In Fig.~\ref{fig:z2_u1} we present wavefunctions 
measured in the PS$_{dir}$ and S$_{alt}$ channels
of GNM$_3$ with either a 
$Z_2$ or a  $U(1)$ chiral symmetry. In both cases the number of fermion species is 
$N_f=4$ and the coupling is $\beta=0.55$. In the $U(1)$ model we fixed the 
vacuum by introducing 
a small fermion bare mass $m_0=0.01$ so that $\langle\sigma\rangle>0$.
The masses  in the $U(1)$ model are: $1.133(3)$ for the S channel, 
$1.123(3)$ for the PS channel 
and $0.567(2)$ 
for the fermion. 
In the $Z_2$ model the masses are $1.085(5)$ for the S channel $1.083(5)$ 
for the PS channel 
and $0.550(2)$ for the fermion.  
These mesons have very broad but quite distinct
wavefunctions, whereas their
masses are almost the same and are very close to twice the fermion mass.
This is further evidence that the wavefunctions 
in this model are more sensitive 
to binding effects than bound state masses.
The difference  between PS and S 
is larger in the $U(1)$ model than in the $Z_2$ model: 
the PS is more tightly bound, 
due to the additional attractive force resulting from 
$\pi$ exchange, which since the $\pi$ is lighter moreover 
also results in a longer-range
$f\bar f$ potential. The kink in the S wavefunction is also 
more pronounced in this case.
Although the various meson masses measured in our simulations are 
in accordance with the large-$N_f$ prediction slightly less than $2M_f$, we believe that 
an accurate measurement of the binding energy can only be extracted from 
simulations on lattices which are bigger than the meson sizes, otherwise
finite size effects can  severely distort the result.

\subsection{Zero Temperature Symmetric Phase}

\begin{figure}[p]

                \centerline{ \epsfysize=3.2in
                             \epsfbox{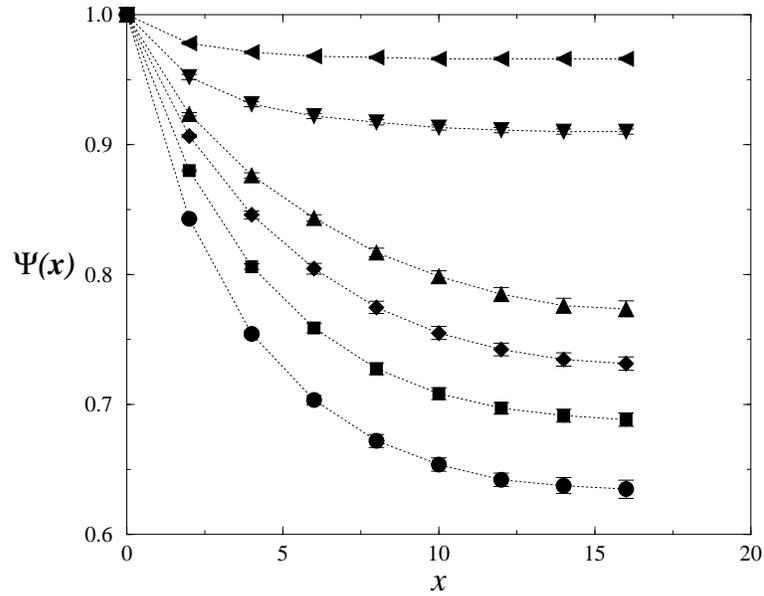}}

\smallskip
\caption[]{PS$_{dir}$ wavefunctions in the broken phase ($\beta=0.45,55,65,0.75$) 
and in the symmetric phase ($\beta=1.0,2.0$); $\beta$ increases 
from the bottom to the top of the figure.}
\label{fig:both_phases}
\end{figure}

We now consider the wavefunctions in the chirally symmetric phase.
As discussed in the Introduction, here there is no dynamically generated 
mass scale,
and no readily identifiable correlation length, although evidence for a
resonance width $\mu$ scaling approximately linearly in ($g_c^2-g^2$) was
reported in \cite{kogut93}. 
In Fig.~\ref{fig:both_phases} 
we show the PS$_{dir}$ wavefunctions of the $Z_2$ 
model in both broken ($\beta<\beta_c$) and symmetric ($\beta>\beta_c$) phases. 
The main observation is that the wavefunction 
becomes broader, i.e. 
the size of the meson increases, as $\beta$ increases,
but that some $f\bar f$ attraction persists beyond the
critical point $\beta_c = 0.85(1)$. 
Counterintuitively, the wavefunction size in lattice units continues to increase
for $\beta>\beta_c$, whereas the correlation length should decrease as we
recede from the critical point. This may indicate that there is no true
bound state in the channel, but merely a positive spatial correlation due to
interparticle attraction, which grows weaker as $\beta\to\infty$.
Before taking this interpretation too seriously, however, it should be remarked
that taken on their own the data of Fig.~\ref{fig:both_phases} would not enable 
the critical coupling to be identified.

\begin{figure}[p]

                \centerline{ \epsfysize=3.2in
                             \epsfbox{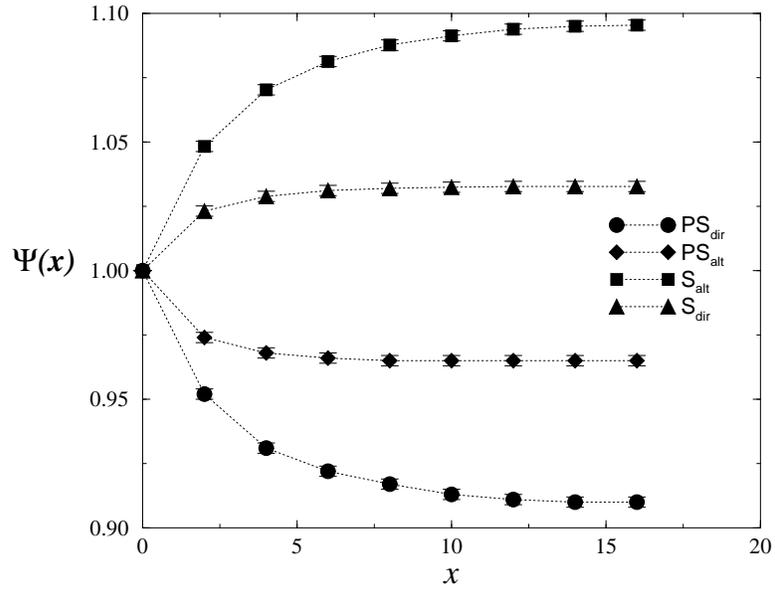}}

\smallskip
\caption[]{S and PS wavefunctions in the zero temperature symmetric phase.}
\label{fig:symmetric}
\end{figure}

In Fig.~\ref{fig:symmetric} we plot data extracted from a simulation on a
$32^3$ lattice at coupling $\beta=1.0$ in the 
symmetric phase in both 
S and PS channels. We generated $10,000$ configurations.
A simple analysis of the correlators $C_M(t)$ 
implies that both are massless. 
Compared with the broken phase, however, the saturation is much
improved, permitting a more complete analysis,
which ultimately permits extraction of both
direct and alternating signals from each channel.
It is possible that 
in this case the wavefunction is yielding information on particle scattering
rather than on an isolated bound state -- this may explain the improved 
saturation, since there is no longer a requirement for the signal from excited
states to die away as $t$ increases before the shape of $\Psi(x)$ stabilises.

We disentangled $\Psi_{dir}(x)$ and $\Psi_{alt}(x)$ by measuring $\Psi(x,t)$,
which is given by
\begin{equation}
\Psi(x,t) = A \Psi_{dir}(x) + (-1)^{t} \tilde{A} \Psi_{alt}(x),
\end{equation}
at large even and odd $t$. The results presented in Fig.~\ref{fig:symmetric}
were extracted from the values of $\Psi$ at $t=14,15$.
A striking feature of the plot is 
that $\vert\Psi_{PS}\vert+\vert\Psi_S\vert$ appears 
independent of $x$. In fact, the meson amplitudes are of opposite signs,
implying that the sum of factors $W_{PS}+W_S$ (see (\ref{eq:wav_latt}))
is of the form $(1-\epsilon(x))$ and hence projects onto odd sink
lattice sites only.
Now, in a chirally symmetric vacuum, meson states built from non-interacting 
fermions (implying a spatially uniform $\Psi$) 
have contributions of precisely this form $G_{oe}G^\dagger_{oe}$; quantum
corrections to $\vert\Psi_{PS}\vert+\vert\Psi_S\vert$ can therefore 
only arise from exchange of an even number of auxiliary bosons, and hence are
suppressed by $O(1/N_f^2)$ at least.
As already observed the PS wavefunctions expand as we increase $\beta$,
with PS$_{dir}$ more tightly bound that PS$_{alt}$.
The similarity in shape may indicate that according to the assignments of
Tab.~\ref{tab:spinflav} both are permitted to scatter in the
$s$-wave.
In contrast to the PS case,
there \emph{is} a qualitative
difference in S channel wavefunctions between broken and symmetric phases.
The wavefunction in both S$_{dir}$ and S$_{alt}$ 
\emph{increases} away from the origin,
implying that close approach of the $f\bar f$ is disfavoured.
This is consistent with the requirement of 
positive parity states to have 
non-zero orbital angular momentum. 

\subsection{Nonzero Temperature}

Next we discuss the effects of non-zero temperature $T$. We have studied
lattices with temporal extent $L_t=16$, for which the critical point for the
Z$_2$ symmetric model is estimated to be $\beta_c^T=0.790(5)$ \cite{strouthos},
and the high temperature chirally restored phase can thus be studied for
$\beta_c^T<\beta<\beta_c^{bulk}$. A particular virtue of GNM$_3$ is that,
unlike unquenched QCD, it
permits a study of hot dynamics with current resources
with $L_t$ sufficiently large to permit
the measurement of thermal masses from correlators in Euclidean time.
We have extracted the thermal fermion mass $M_f^{th}$ by fits to
\begin{equation}
C_f(t)=A(1-(-1)^{t})\sinh(M_f^{th}(t-L_t/2)) + 
B (1+(-1)^{t})\cosh(M_f^{th}(t-L_t/2)).
\label{fermi_prop}
\end{equation}
We performed simulations on $32^2\times 16$ and $96^2\times16$
lattices and we generated approximately  $20,000$ configurations
on the smaller lattice and $35,000$ on the bigger one. 
However, our data are consistent with
the vanishing of $C_f(t)$ for even $t$, corresponding 
to $B=0$ in (\ref{fermi_prop}), signalling a manifest chiral symmetry.
The results for $M_f^{th}$
are given in
Tab.~\ref{tab:mth}.
\begin{table}[ht]
\setlength{\tabcolsep}{1.5pc}
\caption{Thermal fermion masses in the hot phase}
\label{tab:mth}
\begin{tabular*}{\textwidth}{@{}l@{\extracolsep{\fill}}rrrr}
\hline
&$\beta$& $32^2\times16$ & $96^2\times16$ \\
\hline
&0.79& -- &0.048(1)\\
&0.80& 0.044(1)&0.041(1)\\
&0.82& -- &0.034(1)\\
&0.85& 0.026(1)&--\\
\hline
\end{tabular*}
\end{table}
We observe that $M_f$ is reasonably insensitive to the spatial volume of the
lattice.
The masses, though small, are considerably larger than those extracted
from the bulk symmetric phase on a $32^3$ lattice, where typical values found
were $M_f=0.006(1)$ at $\beta=1.0$ and 0.0015(3) at $\beta=2.0$, ie. virtually
massless. The fermion mass thus provides a criterion from distinguishing the two
chirally symmetric phases. In thermal field theory the typical scale for thermal
masses is given by $m^{th}\sim g(T)T$ where $g(T)$ is the coupling strength
associated with a vertex in a perturbative calculation. For GNM$_3$ the
coupling associated with the UV fixed point
is \cite{kogut93}
\begin{equation}
g={4\over{N_f\surd k^2}}.
\end{equation}
For asymptotically high $T$ we thus have $g(T)\propto T^{-1}$ and hence
$\lim_{T\to\infty}M_f^{th}\to \mbox{const}$. For finite $T$ we still expect
$M_f^{th}/T$ to increase more slowly than $T$, implying that for fixed
$L_t$ $M_f^{th}a$ should decrease as $\beta\to\beta_c^{bulk}$, as suggested by
the data of Tab.~\ref{tab:mth}. In this limit the large-$N_f$ expansion suggests
that GNM$_3$ becomes an ideal fermi gas \cite{Gatto}.

\begin{figure}[p]

                \centerline{ \epsfysize=3.2in
                             \epsfbox{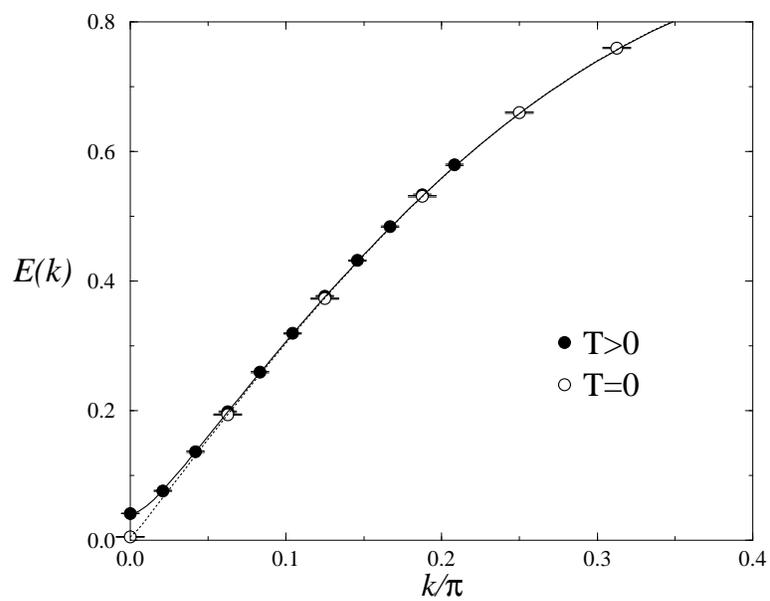}}

\smallskip
\caption[]{Fermion dispersion relations at $T=0$ and $T>0$.}
\label{fig:disperse}
\end{figure}

We have also examined fermion correlation functions at non-vanishing momenta
$\mathbf{k}=2\pi \mathbf{n}/L_s$, $n=0,1,2,\ldots$
and used the energies $E(k)$ extracted from
(\ref{fermi_prop}) to map out the fermion dispersion relation. Results
from both hot and bulk chirally symmetric phases are shown in
Fig.~\ref{fig:disperse}, togther with fits to the lattice form
\begin{equation}
E(k)=A\sinh^{-1}(\sqrt{\sin^2k+M_f^2}).
\end{equation}
The fits yield masses consistent with those extracted by fitting directly
at zero momentum, and in both phases yield values of $A$, related to
the renormalised speed
of light, very close to 1, implying that the principal physical effect of the
hot medium is to generate a non-zero thermal mass. A similar observation
has been made in quenched QCD \cite{Laermann}. The result $A\simeq1$
supports our identification of a non-zero screening mass, or ``pseudogap'',
in studies of the $U(1)$-symmetric model at $T>0$ \cite{babaev}, where
large volume effects precluded using $L_t$ sufficiently large to extract
$E(k)$. Modifications of the low momentum part of the dispersion relation
which is expected 
to have two branches at $T>0$ \cite{weldon} corresponding to 
two kinds of quasiparticle excitations, the
fermion and the hole (known as the plasmino), 
are not visible in our simulations.
Much larger lattices are needed 
to study the structure of the dispersion relation 
at the required low momenta.

\begin{figure}[p]

                \centerline{ \epsfysize=3.2in
                             \epsfbox{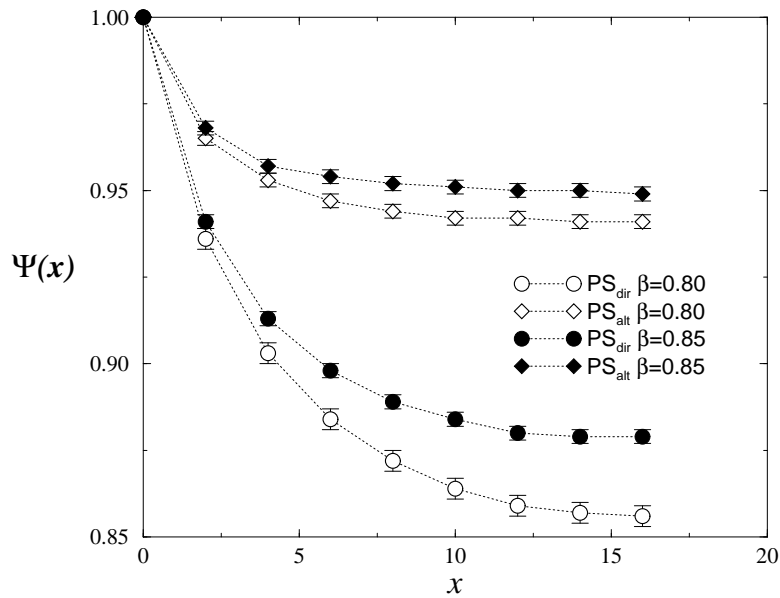}}

\smallskip
\caption[]{PS wavefunctions in the high $T$ symmetric phase.}
\label{fig:t.neq.0}
\end{figure}

As in the bulk symmetric phase, it is necessary to have accurate measurements of
meson masses before wavefunctions can be extracted. On $16\times32^2$ at
$\beta=0.80$ we found a mass 0.075(5) in the PS$_{dir}$ channel, which decreased
to 0.038(3) at $\beta=0.85$, and masses close to if not consistent with zero for
PS$_{alt}$. In the S channel stable signals were found in both
S$_{alt}$ (0.090(2)) and S$_{dir}$ (0.047(2)) at $\beta=0.80$, decreasing to
0.051(5) and 0.037(5) by $\beta=0.85$.
As in the bulk symmetric phase, saturation was found to be quite satisfactory,
which may again be a sign that there is no bound state pole.
The resulting wavefunctions in PS$_{dir}$
and PS$_{alt}$ channels are shown in Fig.~\ref{fig:t.neq.0}. The general shapes
are very similar to those seen in both chirally broken and bulk symmetric
phases, with the width of the wavefunction increasing monotonically with
$\beta$. In principle the $x$ axis should be rescaled with the  correlation
length given by $M_f^{-1}(T=0)$; however, we have not done so
since $\beta=0.85$ is extremely close to the bulk critical point where finite
volume corrections are large, making
an estimate of $M_f(T=0)$ by any other means than naively extrapolating from
smaller $\beta$ using $\nu\simeq1$ impracticable.
Instead, we note that the physical scale of the structure in
Fig.~\ref{fig:t.neq.0} must shrink to zero as $\beta\to\beta_c^{bulk}$, implying
that the region where $f\bar f$ interactions are important becomes pointlike as
$T\to\infty$. This could perhaps be explained in terms of Debye screening of the
$f\bar f$ interaction due to thermal effects. Once again, we are unable to
determine the issue of whether the mesons remain bound states for $T>T_c$
without a detailed study of the spectral function -- the behaviour of
$\Psi(x)$ for large $x$ is subject to large finite volume effects in this
coupling regime.

\section{Summary and Outlook}

We have applied a technique originally
developed for studying mesons in QCD, namely 
analysis of spatial wavefunctions, in a simpler model 
which manifests a variety of interesting
behaviours as both coupling strength and/or temperature are varied. 
We have successfully identified mesonic bound states in the 
chirally broken phase, and exposed the effects of changing the strength
of the interaction, either by altering the number of species $N_f$ or the
chiral symmetry group, much more cleanly than permitted
by a study of the spectrum alone. 
Our results are reminiscent of the well-known property of the variational method
in quantum mechanics that energy eigenvalues are better determined, i.e. less sensitive
to small changes in the variational parameter, than the wavefunctions themselves.
The physical picture is a two-dimensional 
analogue of the quark model, with the $s$-wave nature of the PS state clearly
distinguishable.
Since the states are weakly bound, they have a
larger spatial extent than the corresponding QCD mesons, and hence are more
prone to finite volume corrections.
We have also studied the wavefunctions in two
distinct chirally symmetric phases, a $T=0$ phase for $\beta>\beta_c^{bulk}$,
and a hot phase for $\beta_c^T<\beta<\beta_c^{bulk}$. 
Because of the difficulties in reliably extracting a correlation length 
in either case, our study here has necessarily been more exploratory. 
Nonetheless we have seen evidence for a clear channel dependence 
in the spatial correlation of propagating $f\bar f$ pairs, which may now be due
to a continuum of scattering states rather than isolated bound states.
Interestingly, the clearest distinction between the two symmetric phases
comes from a study of the fermion dispersion relation, which yields a 
non-vanishing thermal mass for the $T>0$ study. If we use our expectations from
the large-$N_f$ approach, then the wavefunctions also provide evidence for 
screening in the hot phase.

The difficulties we have faced with finite volume effects are intimately 
related to the details of the spectrum of the model. In a related 
study \cite{mem} we are analysing the spectral functions of
the model using Maximum Entropy techniques. One advantage is that this approach
permits analysis of the auxiliary fields, corresponding in lattice QCD parlance
to the inclusion of disconnected quark line diagrams as well. It is hoped that 
this complementary approach will also help to resolve the issue of whether 
there are bound states in the chirally symmetric phases.

Finally, another feature of GNM$_3$ is that it permits simulation with a 
non-zero chemical potential $\mu$, making studies of degenerate ``quark matter''
possible \cite{general}. We are currently extending our wavefunction studies
to $\mu\not=0$, where a sharp Fermi surface manifests itself by a  
$\Psi(x)$ which fluctuates in sign, demonstrating 
so-called Friedel oscillations \cite{Tran}.

\section*{Acknowledgements}
Discussions with Chris Allton are greatly appreciated.
SJH and CGS were supported by a Leverhulme Trust grant, and also partially by
EU TMR network ERBFMRX-CT97-0122.
JBK was supported in part by NSF grant  PHY-0102409. 
The computer simulations were done on the Cray SV1's at NERSC, the Cray T90
at NPACI, 
and on the SGI Origin 2000 at the University of Wales Swansea.  


\newpage

\end{document}